\documentclass[preprint,superscriptaddress,amsmath,amssymb,prd,aps,showpacs,floatfix,nofootinbib]{revtex4-1}
\usepackage{graphicx}
\usepackage{dcolumn}
\usepackage{bm}
\usepackage{mathrsfs}
\usepackage{hyperref}
\usepackage{amsmath}
\usepackage{amssymb}
\usepackage{float}
\usepackage{dcolumn}
\usepackage{multirow}

\begin{document}
\title{Molecular picture for the $\Omega(2012)$ revisited}
\date{\today}

\author{Natsumi Ikeno}
\email{ikeno@tottori-u.ac.jp}
\affiliation{Department of Agricultural, Life and Environmental Sciences, Tottori University, Tottori 680-8551, Japan}
\affiliation{Departamento de F\'{\i}sica Te\'orica and IFIC, Centro Mixto Universidad de Valencia-CSIC, Institutos de Investigaci\'on de Paterna, Aptdo.22085, 46071 Valencia, Spain}

\author{Genaro Toledo} 
\email{toledo@fisica.unam.mx}
\affiliation{Departamento de F\'{\i}sica Te\'orica and IFIC, Centro Mixto Universidad de Valencia-CSIC, Institutos de Investigaci\'on de Paterna, Aptdo.22085, 46071 Valencia, Spain}
\affiliation{ Instituto de F\'{\i}sica, Universidad Nacional Aut\'onoma de M\'exico, A.P. 
20-364, Ciudad de M\'exico 01000, M\'exico.}

\author{Eulogio Oset}
\email{oset@ific.uv.es}
\affiliation{Departamento de F\'{\i}sica Te\'orica and IFIC, Centro Mixto Universidad de Valencia-CSIC, Institutos de Investigaci\'on de Paterna, Aptdo.22085, 46071 Valencia, Spain}

\begin{abstract}
We conduct a study of the interaction of the $\bar K \Xi^*$, $\eta \Omega$($s$-wave) and $\bar K \Xi$($d$-wave) channels within a coupled channel unitary approach where the transition potential between the $\bar K \Xi^*$ and $\eta \Omega$ channels is obtained from chiral Lagrangians. The transition potential between $\bar K \Xi^*$, $\eta \Omega$ and $\bar K \Xi$ is taken in terms of free parameters, which together with a cut off to regularize the meson-baryon loops are fitted to the $\Omega(2012)$ data. We find that all data including the recent Belle experiment on $\Gamma_{\Omega^* \to \pi \bar K \Xi}/ \Gamma_{\Omega^* \to \bar K \Xi}$, are compatible with the molecular picture stemming from meson baryon interaction of these channels.
\end{abstract}

\maketitle


\section{Introduction}

  The discovery of the $\Omega(2012)$ state by the Belle collaboration~\cite{belle1}
prompted much theoretical work on the issue, with pictures inspired by quark models and also molecular pictures based on the meson-baryon interaction. $\Omega$ excited states appear naturally in different versions of quark models, nonrelativistic quark models~\cite{isgur,chen,anmet,anzou,hayne,roberts}, relativistic quark models~\cite{capstick,metsch,faustov}. Different QCD based approaches have also been used to get spectra, as QCD sum rules~\cite{aliev,azizi}, the Skyrme model~\cite{oh}, large $N_c$ considerations~\cite{goity} and lattice QCD calculations \cite{engle}. Methods based on symmetries, as SU(3)~\cite{polyakov}, and the algebraic method~\cite{iachelo} have also contributed to this topic. In those models the $\Omega(2012)$ approximately fits as a $J^P=3/2^-$ state, yet there is a dispersion of the results for the masses for this state that ranges from 1978 MeV in Ref.~\cite{oh} to 2049 MeV in Ref.~\cite{engle}. In addition, models that incorporate five quark components reduce the mass of the state in about 200 MeV~\cite{yuan,anmet,anzou}.

   Some low lying $3/2^-$ states are also obtained as molecular states from the interaction of pseudoscalar mesons and ground state baryons of ($3/2^+$)~\cite{kolo,sarkar,xieome}. Concretely, for the generation of the $\Omega(2012)$ state the interacting channels are $\bar K \Xi^*$ and $\eta \Omega$, and the interaction is obtained from chiral Lagrangians unitarizing in coupled channels. The interaction in this case is peculiar since the diagonal potential of the two channels is null, and only the transition potential between these two channels is non zero. This has as a consequence that the mass of the generated bound $\Omega$ state depends more strongly on the parameters of the model than in other ordinary molecular states. In this sense the data of Ref.~\cite{belle1} was most welcome since it provides the information needed to tune the parameters of the theory and come out with more accurate predictions. Thus, after the Belle experiment reported the $\Omega(2012)$ state, some works were devoted to trying to understand the nature of this state. In this sense, in Ref.~\cite{pavon} the two channels $\bar K \Xi^*$ and $\eta \Omega$ were considered with the input of the chiral Lagrangians, and the parameters of the model, related to the form factors that regularize the meson baryon loops, were tuned to the experiment. The $\bar K \Xi^*$ channel is bound by about 11 MeV, but due to the $\Xi^*(1530)$ width there is a decay into $\bar K \pi \Xi$. In addition, a model is made for the $\bar K \Xi^* \to \bar K \Xi$ transition and it is concluded that the $\Omega(2012)$ width comes approximately with the same strength from the $\bar K \pi \Xi$ and $\bar K \Xi$ decay channels. In Ref.~\cite{bszou} the $\Omega(2012)$ state is supposed to be a $\bar K \Xi^*$ bound state and the coupling of the state to the $\bar K \Xi^*$  channel is obtained using the Weinberg compositeness condition \cite{weinberg,baru,danijuan,thomas}. The transition from this state to the $\bar K \Xi$ channel is evaluated by means of a triangle diagram that involves vector exchange, and the strength of the transition is small, such that practically all the decay of the $\Omega(2012)$ state goes to $\bar K \pi \Xi$. Actually, vector exchange through this mechanism was found to be dynamically very suppressed in Refs.~\cite{xiegeng} and \cite{pavao}. In Refs.~\cite{xiegeng} and \cite{pavao} the chiral unitary approach was used, and in Ref.~\cite{xiegeng} the transition $\bar K \Xi^* \to \bar K \Xi$ is evaluated by means of a triangle diagram that involves baryon exchange. By the contrary, in Ref.~\cite{pavao}, given the difficulty to evaluate the $\bar K \Xi^* \to \bar K \Xi$ transition, which proceeds in $d$-wave, the matrix elements for $\bar K \Xi^* \to \bar K \Xi$ and $\eta \Omega \to \bar K \Xi$ transitions are parametrized and fit to the data. In addition, the coupled channels were extended to include explicitly the $\bar K \Xi$ channel. In both cases a qualitative agreement with the data was found.  In Ref.~\cite{pavao} some fits to the mass and width of the $\Omega(2012)$ state were done and the partial decay widths to $\bar K \pi \Xi$, $\bar K \Xi$ were found to be of similar strength, with the $\bar K \Xi$ decay channel dominant. 

   This was the situation until the Belle collaboration presented the results from an experiment~\cite{belle2} that showed that the ratio of the $\bar K \pi \Xi$ width to the $\bar K \Xi$ width is smaller than 11.9\%. This paper concludes ``Our result strongly disfavors the molecular interpretation of Ref.~\cite{bszou} and is in tension with the predictions
of Refs.~\cite{pavon,xiegeng,pavao,polyakov}, also based on molecular interpretations''. 

   There has already been a feedback to this work and in Ref.~\cite{linzou} the authors redo the analysis of Ref.~\cite{bszou} concluding that the new data favor a molecular $\bar K \Xi^*$ state but in $p$-wave rather than $s$-wave. In Ref.~\cite{qifang} a nonrelativistic quark model is used and the $\Omega(2012)$ state is identified with a $3/2^-$ ($1^2 P_{3/2}$) state. The strong decay width is also evaluated and a width of 5.6 MeV is obtained in the $\bar K \Xi$ channel and zero in the $\bar K \Xi^*$ one. In Ref.~\cite{gulyu} the molecular picture is retaken using the Weinberg compositeness condition, but contrary to Ref.~\cite{bszou}, the molecule is considered not from the $\bar K \Xi^*$ state but from a coherent mixture of $\bar K \Xi^*$ and $\eta \Omega$. The work shares with Ref.~\cite{pavao} and the present work the relevance of the $\eta \Omega$ component to stabilize the molecular state. With large uncertainties, depending on the parameters used, they find widths of the order of magnitude of the experiment, where the $\bar K \Xi$ decay channel is dominant, but the $\bar K \pi \Xi$ decay channel has also a relatively large strength and a ratio of this decay width to the $\bar K \Xi$ width smaller than 12\% is not easy to get, although possible within theoretical uncertainties. The conclusion of the work is that ``The prediction given here can hopefully support a possible structure interpretation of the $\Omega(2012)$''.

    There is, however, a detail that has passed unnoticed in the former works and this is the cut made in Ref.~\cite{belle2} to conclude that the detected $\bar K \pi \Xi$ state comes from  $\bar K \Xi^*$. For this purpose a cut is made in the invariant mass of $\pi \Xi$, demanding  
   
\begin{equation}
 1.49~{\rm GeV} < M_{\rm inv}(\pi \Xi) < 1.53~{\rm GeV}
\label{eq:belle.cut}
\end{equation}

  This cut was not implemented in Refs.~\cite{pavon,xiegeng,pavao}, hence a proper comparison demands to redo the calculations implementing this cut. On the other hand the new information of Ref.~\cite{belle2} is very valuable to further pin down the unknown parameters of the theory. With this double perspective we take the task to reanalyze the work of Ref.~\cite{pavao} in order to see to which extend the data rule out the molecular picture or not. 
 Anticipating the results, we find that the results are compatible with the molecular picture in coupled channels, but in passing we will learn more about the role played by the $\eta \Omega$ channel in this problem.

\section{Formalism}\label{formalism}
We follow the steps of Ref.~\cite{pavao} and take the coupled channels $\bar K \Xi^*$, $\eta \Omega$, $\bar K \Xi$. The first two channels are in $s$-wave and the latter one in $d$-wave. The $3 \times 3$ scattering matrix calculated with the Bethe Salpeter equation is given by 
\begin{equation}
T =\left[1-VG\right]^{-1} V,
\label{eq:BS}
\end{equation}
where the transition potential is given by
\begin{equation}
\label{eq:Vm}
V= \begin{matrix}
 \begin{matrix}
\bar{K} \Xi^* & \eta \Omega &  \bar{K}\Xi
\end{matrix}& \\ 
\begin{pmatrix}
 0& 3F & \alpha q^2_{\rm on}\\ 
 3F& 0 & \beta q^2_{\rm on}\\ 
 \alpha q^2_{\rm on} &  \beta q^2_{\rm on} & 0 
\end{pmatrix} & \begin{matrix}
\bar{K} \Xi^* \\ 
\eta \Omega \\ 
\bar{K}\Xi
\end{matrix}
\end{matrix}
\end{equation}
with
\begin{equation}
 F = -\frac{1}{4f^2} (k^0 +k'^0) ; \hspace{5mm} q_{\rm on}=  \frac{\lambda^{1/2}\left(s,m_{\bar K}^2,m_{\Xi}^2 \right)}{2 \sqrt{s}},
 \end{equation}
with 
$f=93$ MeV, the pion decay constant, and $k^0$, $k'^0$ the energies of initial and final mesons, respectively.
In Eq.~(\ref{eq:Vm}) the transition potentials between $\bar K \Xi^*$ and $\eta \Omega$ are taken from the chiral Lagrangians~\cite{sarkar}, while the transition potential between $\bar K \Xi$ and $\bar K \Xi^*$ or $\eta \Omega$, which proceed in $d$-wave, are taken in terms of the free parameters $\alpha$, $\beta$. The potential of Eq.~(\ref{eq:Vm}) should have $\bm q^2$, instead of $q_{\rm on}^2$, when it is used inside loops, but technically it is more practical to include this dependence in the meson baryon loop function $G$ of Eq.~(\ref{eq:BS}), which is given by 
\begin{equation}
\label{eq:loopmat}
G(\sqrt{s})=\begin{pmatrix}
G_{\bar{K}\Xi^*}(\sqrt{s}) & 0 & 0\\ 
0 & G_{\eta \Omega}(\sqrt{s}) &0 \\ 
0 & 0 & G_{\bar{K}\Xi}(\sqrt{s})
\end{pmatrix},
\end{equation}
where
\begin{equation}
\label{eq:loop}
G_i(\sqrt{s}) = \int_{|\bm{q}|<q_{\text{max}}} \frac{d^3 q}{(2 \pi)^3} \frac{1}{2 \omega_i(\bm{q})} \frac{M_i}{E_i(\bm{q})} \frac{1}{\sqrt{s}-\omega_i(\bm{q})-E_i(\bm{q}) + i \epsilon} ,
\end{equation}
for $i= \bar K \Xi^*, \eta \Omega$, with $\omega_i(\bm{q}) = \sqrt{m_i^2 + \bm q^2}$, $E_i (\bm{q}) = \sqrt{M_i^2 + \bm q^2}$ and $m_i$, $M_i$ the meson and baryon masses of the $i$ channels. For the $d$-wave channel, $\bar K \Xi$, the $G$ function is then given by
\begin{equation}
G_{\bar{K}\Xi}(\sqrt{s}) = \int_{|\bm{q}|<q'_{\text{max}}} \frac{d^3 q}{(2 \pi)^3} \frac{ (q/q_{\rm on})^4}{2 \omega_{\bar{K}}(\bm{q})} \frac{M_{\Xi}}{E_{\Xi}(\bm{q})} \frac{1}{\sqrt{s}-\omega_{\bar{K}}(\bm{q})-E_{\Xi}(\bm{q}) + i \epsilon},
\label{eq:loop2}
\end{equation}

In Eqs.(\ref{eq:loop})(\ref{eq:loop2}), $q_{\rm max}$, $q'_{\rm max}$ are the cut off to regularize the loop functions. We take them equal, around 700~MeV and do the fine tuning of this parameter to the experimental data. The $\bar K \Xi$ channel is a weak channel in the interaction, but provides the main source for the decay. Due to this, the diagonal $\bar K \Xi \to \bar K \Xi$ transition is taken zero as in Ref.~\cite{pavao}.

The mixing of $s$-waves and $d$-waves is not common in works of the chiral unitary approach, where only $s$-wave channels are normally used, but it is not new. Note that since the $d$-wave only appears in the transition from $s$-wave to $d$-wave in a loop of $\bar K \Xi$ connected to two external $s$-wave channels the $d$-wave of the two vertices gives rise to an $s$-wave upon the $q$ integration, producing an additional contribution to the $s$-wave transition between the $\bar K \Xi^*$ or $\eta \Omega$ states.  This formalism was already used in the study of the $\Lambda(1520)$ in Refs.~\cite{sarkarmano,sarcarora}, where the $\pi \Sigma^*$ and $K \Xi^*$ channels are in $s$-wave and the $\bar K N$ and $\pi \Sigma$ in $d$-wave.  In Ref.~\cite{sarkarmano}, the $\Lambda(1520) \to \pi^0 \pi^0 \Lambda$ reaction was studied with this formalism and good agreement with data was obtained. Similarly, in Ref.~\cite{sarcarora} the $K^- p \to \Lambda \pi \pi$, $\gamma p \to K^+ K^- p$, $\gamma p \to K^+ \pi^0 \pi^0 \Lambda$ and $\pi^- p \to K^0 K^- p$ reactions were studied with this formalism and again a fair agreement with data was found. The formalism was also used to study the radiative decay of the $\Lambda(1520)$ to $\gamma \Sigma$ and $\gamma \Lambda$ in Ref.\cite{misha},  where a discussion on possible other components was made.

\begin{figure}[tb!]
  \centering
  \includegraphics[width=0.5\textwidth]{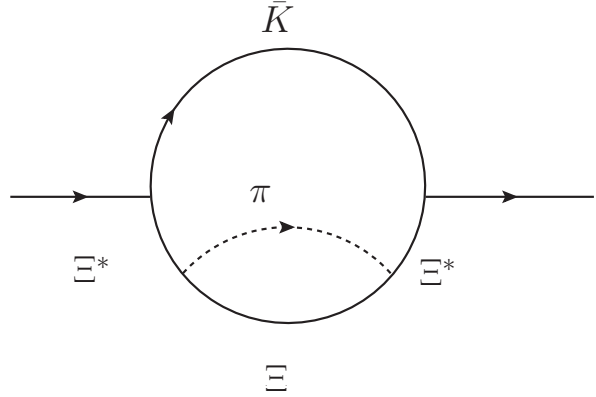}
  \caption{$G_{\bar K \Xi^*}$ function accounting for the $\Xi^* \to \pi \Xi$ decay width.}
  \label{fig:1}
\end{figure}

Since we are close to the $\bar K \Xi^*$ threshold, it is important to take into account the mass distribution of the $\Xi^*$, due to its width for $\Xi^* \to \pi \Xi$ decay, and the $G_{\bar K \Xi^*}$ is convolved with the $\Xi^*$ mass distribution (see technical details in Ref.~\cite{pavao}). It is interesting to note that by doing this, one is including the new diagram of Fig.~\ref{fig:1} in the $G_{\bar K \Xi^*}$ function, as a consequence of which, $G_{\bar K \Xi^*}$ gets an imaginary part when the $\bar K \pi \Xi$ state is placed on shell in the loop, in other words, now one is accounting for the $\Omega(2012) \to \pi \bar K \Xi$ decay in the coupled channels approach. Comparison of the $\Omega(2012)$ with convolution, which accounts for $\bar K \Xi$ and $\pi \bar K \Xi$ decays, and without convolution, which accounts only for the $\bar K \Xi$ decay, gives us one estimate of the $\Omega(2012)$ decay width into the $\bar K \Xi^* \to \bar K \pi \Xi$ decay channel, the one measured in Ref.~\cite{belle2}.

The couplings $g_i$ of the resonance $\Omega(2012)$ to the different channels are obtained from the residue of the $T$-matrix of the pole in the second Riemann sheet (see Ref.~\cite{pavao} for details). Close to the pole we have
\begin{equation}
\label{gl}
T_{ij} = \frac{g_i g_j}{z -z_R}\,  (z, {\rm complex~energy;~} z_R, {\rm complex~ pole~ position}),
\end{equation}
\begin{equation}
g_i^2 =\lim_{z \to z_R}(z - z_R) T_{ii}; 
~~ g_j = g_i \frac{T_{ij}}{T_{ii}}|_{z=z_R}.
\label{eq:g_sing}
\end{equation}

We take the first of the former equations to determine $g^2_{\bar K \Xi^*}$ and $g_{\bar K \Xi^*}$ has then an arbitrary sign but the second equation of Eq.~(\ref{eq:g_sing}) allows to get the relative sign of the others couplings with respect to $g_{\bar K \Xi^*}$.

The convolution of $G_{\bar K \Xi^*}$ introduces some changes, and if the state obtained is close to the $\bar K \Xi^*$ threshold the neat pole obtained without the convolution can give rise to a pole distribution (a cut), as discussed in Ref.~\cite{garzon}, in which case, in order to compare with the empirical amplitude
\begin{equation}
T_{ii} = \frac{g_i^2}{\sqrt{s}-M_R+i \frac{\Gamma}{2}},
\end{equation}
we take the $T_{ii}$ matrix at its peak and get,
\begin{equation}
 |g_{i}|^2 = \frac{\Gamma}{2} \sqrt{|T_{ii}|^2_{\rm max}},
\label{eq:coupl.g2}
\end{equation}
and $\Gamma$ is also obtained from the shape of $|T|^2$, from
\begin{equation}
 \frac{\Gamma}{2} = \sqrt{s} - M_R,
\end{equation}
for $\sqrt{s}$ where $|T|^2(\sqrt{s})/|T|^2_{\rm max}$ is $1/2$, with $M_R$ the value of $\sqrt{s}$ at the peak of $|T|^2$.

\begin{figure}[tb!]
  \centering
  \includegraphics[width=0.50\textwidth]{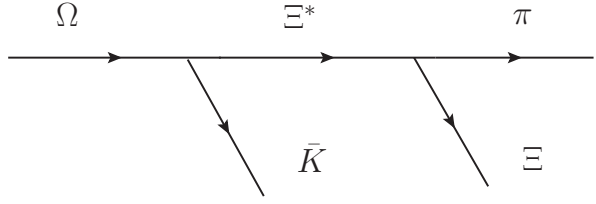}
  \caption{Decay diagram of $\Omega(2012)$ into $\bar K \Xi^*$, with $\Xi^* \to \pi \Xi$.}
  \label{fig:2}
\end{figure}

In order to evaluate the ratio of $\Gamma(\pi \bar K \Xi)/\Gamma_{\bar K \Xi}$ with the cuts of the experiment and compare with the experimental data~\cite{belle2}, we perform the explicit calculation of the process depicted in Fig.~\ref{fig:2}

In order to evaluate the amplitude for the process of Fig.~\ref{fig:2} we need the coupling $\Omega \to \bar K \Xi^*$, which is evaluated above, and the $\Xi^* \to \pi \Xi$ coupling. It suffices to consider that this is a $p$-wave coupling, proportional to $\tilde p_\pi$ in the $\Xi^*$ rest frame, and then we have
\begin{equation}
 t_{\Omega, \pi \bar K \Xi} = g_{\Omega, \bar K \Xi^*}
 \frac{1}{M_{\rm inv}(\pi \Xi) -M_{\Xi^*} +i \frac{\Gamma_{\Xi^*}}{2}} g_{\Xi^*, \pi \Xi}  \   \tilde p_\pi,
\end{equation}
with
\begin{equation}
 \tilde p_\pi = \frac{\lambda^{1/2}\left(M_{\rm inv}^2(\pi \Xi),m_{\pi}^2,M_{\Xi}^2 \right)}{2 M_{\rm inv}(\pi \Xi)},
\label{eq:Ppi}
\end{equation}
with $g_{\Xi^*, \pi \Xi}$ given in terms of $\Gamma_{\Xi^*}$ via
\begin{equation}
 \Gamma_{\Xi^*} = \frac{1}{2\pi}\frac{M_{\Xi}}{M_{\Xi^*}} g^2_{\Xi^*, \pi \Xi} \  \tilde p^3_\pi,
\label{eq:GamXis}
\end{equation}
taking $\Gamma_{\Xi^*, {\rm on}} = 9.1 \pm 0.5$~MeV~\cite{pdg}.
Then we have for the $\pi \Xi$ mass distribution
\begin{equation}
 \frac{d \Gamma_{\Omega}}{ d M_{\rm inv}(\pi \Xi)} = \frac{1}{(2\pi)^3} \frac{2 M_{\Xi} \ 2M_{\Omega(2012)} }{4 M_{\Omega(2012)}^2} \ p_{\bar K} \ \tilde p_{\pi} |t_{\Omega, \pi \bar K \Xi}|^2,
\label{eq:dGdM}
\end{equation}
 with 
\begin{equation}
 p_{\bar K} =  \frac{\lambda^{1/2} (M^2_{\Omega(2012)},m_{\bar K}^2,M^2_{\rm inv}(\pi \Xi) )}{2 M_{\Omega(2012)}}.
\end{equation}

The way to evaluate the width for $\pi \bar K \Xi$ decay described above is new with respect to the one used in Ref.~\cite{pavao}, which was based on the comparison of the $\Omega(2012)$ widths with and without convolution of the $\bar K \Xi^*$ channel. Here we will compare both methods, but the new one is needed to implement the cuts of the experiment of Ref.~\cite{belle2}. There is also another novelty with respect to Ref.~\cite{pavao}, since both in the convolution of $G_{\bar K \Xi^*}$ and in the evaluation of $d \Gamma_{\Omega} / dM_{\rm  inv}(\pi \Xi)$ of Eq.~(\ref{eq:dGdM}) we take the $\Xi^*$ width energy dependent. This is, we take
\begin{equation}
 \Gamma_{\Xi^*} =\Gamma_{\Xi^*, \rm{on}} \ \frac{\tilde p^3_\pi}{\tilde p^3_{\pi,\rm{on}}} \ \theta(M_{\rm inv}(\pi \Xi) - m_\pi -M_\Xi),
\label{eq:Gamd.Xis.Edep}
\end{equation}
with $\tilde p_\pi$, $\tilde p_{\pi, {\rm on}}$ given by Eq.~(\ref{eq:Ppi}) with $M_{\rm inv} (\pi \Xi)$ for $\tilde p_\pi$ and $M_{\rm inv} (\pi \Xi) = M_{\Xi^*}$ for $\tilde p_{\pi, {\rm on}}$. By using explicitly Eq.~(\ref{eq:GamXis}) we avoid having to evaluate $g_{\Xi^*, \pi \Xi}$ and we directly obtain
\begin{equation}
 \frac{d \Gamma_{\Omega}}{ d M_{\rm inv}(\pi \Xi)} = \frac{1}{(2\pi)^2} \frac{M_{\Xi^*}}{M_{\Omega(2012)}} \ p_{\bar K} \ g^2_{\Omega, \bar K \Xi^*}
\frac{\Gamma_{\Xi^*}}{\left| M_{\rm inv}(\pi \Xi) -M_{\Xi^*} +i \frac{\Gamma_{\Xi^*}}{2}\right|^2}. 
\label{eq:dGamdM2}
\end{equation}
One can see that in the limit of $\Gamma_{\Xi^*} \to 0$, 
$( {\rm Im}(M_{\rm inv}(\pi \Xi)  -M_{\Xi^*}  +i \frac{\Gamma_{\Xi^*}}{2} )^{-1} \to -\pi \delta(M_{\rm inv}(\pi \Xi)  -M_{\Xi^*} ) )$
one obtains the ordinary decay formula for $\Omega(2012) \to \bar K \Xi^*$, indicating that the normalization has been correctly taken into account.

\section{Results}\label{result}
\begin{figure}[tb]
\centering.
 \includegraphics[width=0.80\textwidth]{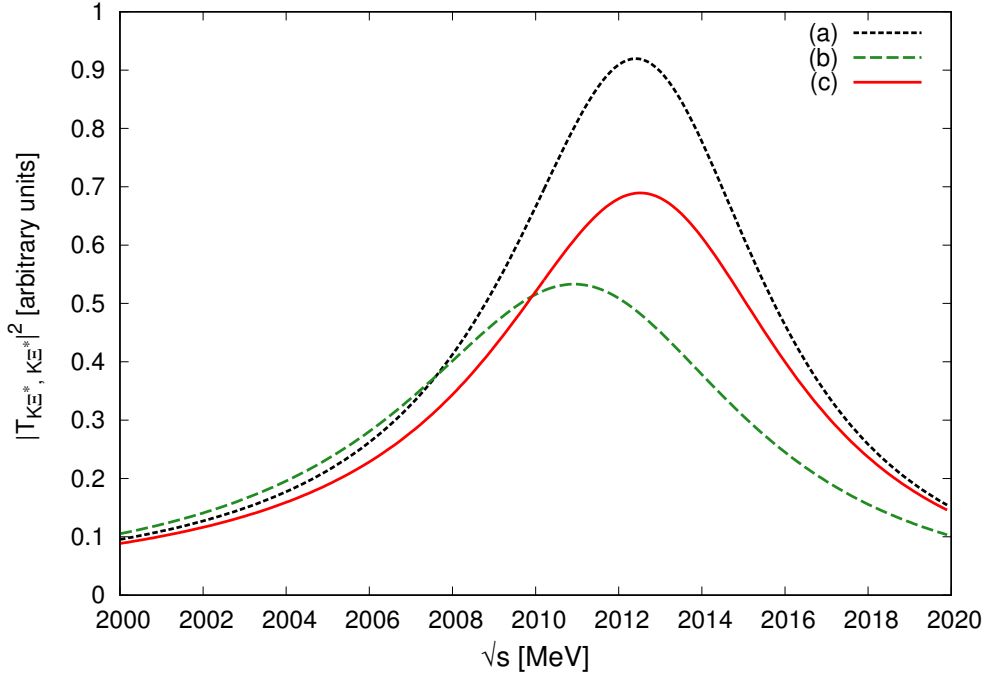}
\caption{$|T|^2$ for the diagonal $\bar K \Xi^*$ channel with three options: (a) $G_{\bar K \Xi^*}$ without convolution; (b) $G_{\bar K \Xi^*}$ with convolution and $\Gamma_{\Xi^*}$ fixed; (c) $G_{\bar K \Xi^*}$ with convolution and $\Gamma_{\Xi^*}$ energy dependent.}
\label{fig:3}
\end{figure}

 We shall make a fit to the experimental data by varying the 3 parameters $q_{\rm max}$, $\alpha$, $\beta$. The experimental data are 
\begin{equation}
 M_{\Omega(2012)} = 2012.4 \pm 0.7 \pm 0.6 ~{\rm MeV},
\label{eq:Exp.mass}
\end{equation}
\begin{equation}
 \Gamma_{\Omega(2012)} = 6.4^{+2.5}_{-2.0} \pm 1.6  ~{\rm MeV},
\label{eq:Exp.width}
\end{equation}
\begin{equation}
\frac{\Gamma_{\Omega}(\pi \bar K \Xi)}{\Gamma_{\Omega,\bar K \Xi}} < 11.9 ~\%.
\label{eq:Exp.ratio}
\end{equation}
We find a reasonable compromise with the parameters
\begin{equation}
 q_{\rm max} = 704~{\rm MeV}; ~~ \alpha =4.5 \times 10^{-8}~{\rm MeV}^{-3};~~
 \beta =5.1 \times 10^{-8}~{\rm MeV}^{-3}  ,
\end{equation}
with $q_{\rm max}$ a little smaller than in Ref.~\cite{pavao} (735~MeV) and the weight of $\beta$ with respect to $\alpha$ also bigger than in Ref.~\cite{pavao}. 
In Fig.~\ref{fig:3}, we show $|T|^2$ for the diagonal $\bar K \Xi^*$ channel in three options; (a) with $G_{\bar K \Xi^*}$ no convolved, 
(b) with $G_{\bar K \Xi^*}$ convolved with $\Gamma_{\Xi^*}$ energy independent fixed to the nominal $\Xi^*$ width (the option of Ref.~\cite{pavao}) and (c) $\Gamma_{\Xi^*}$ energy dependent, Eq.~(\ref{eq:Gamd.Xis.Edep}), the present option.

We observe that options (a), (c) give a similar mass, but the mass of option (b) is a bit displaced to lower values. We find (we refer by $\Omega^*$ the $\Omega(2012)$ for simplicity),
\begin{eqnarray}
&{\rm (a)}&~M_{\Omega^*, {\rm non}} = 2012.4 ~{\rm MeV}, \nonumber\\
&{\rm (b)}&~M_{\Omega^*, {\rm con (Eind)}} = 2011.0 ~{\rm MeV}, \\
&{\rm (c)}&~M_{\Omega^*, {\rm con (Edep)}} = 2012.2 ~{\rm MeV}. \nonumber
\end{eqnarray}
From the value of $\sqrt{s}$ at mid strength of $|T|^2$, we find the widths
\begin{eqnarray} \label{eq:results.Gam}
&{\rm (a)}&~\Gamma_{\Omega^*, {\rm non}} = 7.6~{\rm MeV}, \nonumber\\
&{\rm (b)}&~\Gamma_{\Omega^*, {\rm con (Eind)}} = 9.9 ~{\rm MeV}, \\
&{\rm (c)}&~\Gamma_{\Omega^*, {\rm con (Edep)}} = 8.6 ~{\rm MeV}. \nonumber
\end{eqnarray}
The numbers for the mass and width for the case with energy dependent convolution are compatible with the experimental numbers within errors.
From these, by taking
\begin{eqnarray}
\frac{\Gamma_{\Omega^*, {\rm con (Edep)}} - \Gamma_{\Omega^*, {\rm non}}}{\Gamma_{\Omega^*, {\rm non}}} = 13.1~\%,
\label{eq:res.ratio}
\end{eqnarray}
we can compare with the experimental ratio of Eq.~(\ref{eq:Exp.ratio}). We see that the ratio obtained is a bit larger than the 11.9~\% experimental upper bound, but the cuts on $M_{\rm inv}(\pi \Xi)$ of the experiment are not implemented there. We must comment that it is difficult to obtain smaller ratios of Eq.~(\ref{eq:res.ratio}) by changing the parameters, while still being in agreement with the mass and width of the $\Omega^*$ resonance. In any case, Eq.~(\ref{eq:res.ratio}) improves the agreement with data compared to the result in Ref.~\cite{pavao}(around 90~\%). It is interesting to see that the use of the convolution with the energy dependent $\Xi^*$ width improved this ratio. Indeed, if we use Eq.~(\ref{eq:res.ratio}) with $\Gamma_{\Omega^*, {\rm con (Eind)}}$ the ratio becomes 30~\%, which is bigger than before. The realistic consideration of the $\Xi^{*}$ width of Eq.~(\ref{eq:Gamd.Xis.Edep}) is a factor that renders the fraction of Eq.~(\ref{eq:res.ratio}) smaller.

Next we look into the couplings. In the case of no convolution, we find the pole at 
\begin{equation}
{\rm (a)} ~M_{\Omega^*} ({\rm pole}) = (2013.0,~ i 4.6) ~{\rm MeV}, \nonumber
\end{equation}
with a mass very similar to the one quoted above from the peak of $|T|^2$, and the width, 2~Im$M_{\Omega^*}=9.2$~MeV, also close to the one quoted before.
From the residues of $T_{ij}$ at the pole, we obtain the couplings shown in Table~\ref{table:1}. There we also show the wave function at the origin for the $s$-wave states, $g_i G_i$, calculated at the peak~\cite{danijuan}, and the probability of each channel $-g_{i}^2 \frac{\partial G}{\partial \sqrt{s}}$.
For a dynamically generated state, $ \displaystyle \sum_{i} (-)g_{i}^2 \frac{\partial G}{\partial \sqrt{s}} = 1$, and in the case of only bound channels, each term of this sum is the probability of the respective channel~\cite{danijuan}. In the present case, the real part of each term can approximately be considered the probability of this channel for the closed ones.

\begin{table}[tb!]
\caption{\label{table:1} 
Couplings $g_i$, wave functions at the origin $g_i G_i$(MeV), and probabilities $-g_{i}^2 \frac{\partial G_i}{\partial \sqrt{s}}$ obtained from the pole. $M_{\Omega^*}=2013.0$~MeV. The couplings $\tilde g_i$ are obtained from $|T|^2$ using Eq.~(\ref{eq:coupl.g2}). The couplings $g_{i, {\rm conv}}$ are obtained from $|T|^2$ using Eq.~(\ref{eq:coupl.g2}) with the $G_{\bar K \Xi^*}$ convolved with the energy dependent $\Xi^*$ width. In parenthesis, we write the threshold mass of each channel in MeV units.
}
\begin{tabular}{cccc}
\hline\hline
& $\bar{K} \Xi^*$ (2027) & $\eta \Omega $ (2220) & $\bar{K} \Xi$ (1812)  \\
\hline
$g_i$ & $2.03 - i0.15$  &~~~$2.93-i0.16$ &~~~$-0.45 +i0.02$  \\
$|\tilde g_i|$ & 1.91   & 2.76  & 0.42  \\
$|g_{i, {\rm conv}}|$ & 1.89  & 2.73 & 0.42 \\
wf$_{ i}$($g_i G_i$) &~~~$-35.46 + i1.09 $  &~~~$-23.78 + i 1.07$  & $-$  \\
$-g_{i}^2 \frac{\partial G_i}{\partial \sqrt{s}}$ &~~$0.68 + i0.04 $ &~~$0.17 - i0.01 $  & $-$  \\
\hline\hline
\end{tabular}
\end{table}

Alternatively we can also obtain the width into the $\bar K \Xi$ channel using the formula,
\begin{equation}
\Gamma= \frac{1}{2 \pi} \frac{M_{\Xi}}{M_{\Omega^*}} \ g_{\bar K \Xi}^2 \ p_{\bar K},
\end{equation}
with $p_{\bar K}$ the $\bar K$ momentum in the $\Omega^*$ rest frame, using the value of $g_{i, {\rm conv}}$ of Table~\ref{table:1}, which gives us 7.4~MeV, in agreement with the 7.6 MeV of $\Gamma_{\Omega^* ,{\rm non}}$ of Eq.~(\ref{eq:results.Gam}). Note that by using Eq.~(\ref{gl}) for all the channels, in the $d$-wave case the $\bm{q}^2$ factor is implicitly included in the coupling $g$.
We can see that the couplings obtained are very similar to those in Ref.~\cite{pavao}, but the strength of $\eta \Omega$ and $\bar K \Xi$ are a bit bigger. We also can see that the strength of the wave function at the origin, as well as the probability, dominates for the $\bar K \Xi^*$ state. 
Since this is the magnitude that enters the evaluation of short range observables, one can conclude that the $\bar K \Xi^*$ state is the dominant component in the $\Omega^*$ wave function. Note, however, that the inclusion of the $\eta \Omega$ channel is what has made the appearance of the $\Omega^*$ bound state possible since the diagonal potential of the $\bar K \Xi^*$ channel is null and hence cannot produce any bound state by itself. As a consequence, it is not surprising that both the strength of the couplings and the wave functions at the origin are of the same size.  
We also observe that the value of the couplings extracted from the residues of the pole and those of $|\tilde g_i|$ from $|T|^2$ via Eq.~~(\ref{eq:coupl.g2}) are quite similar.

\begin{figure}[tb!]
\centering.
\includegraphics[width=0.80\textwidth]{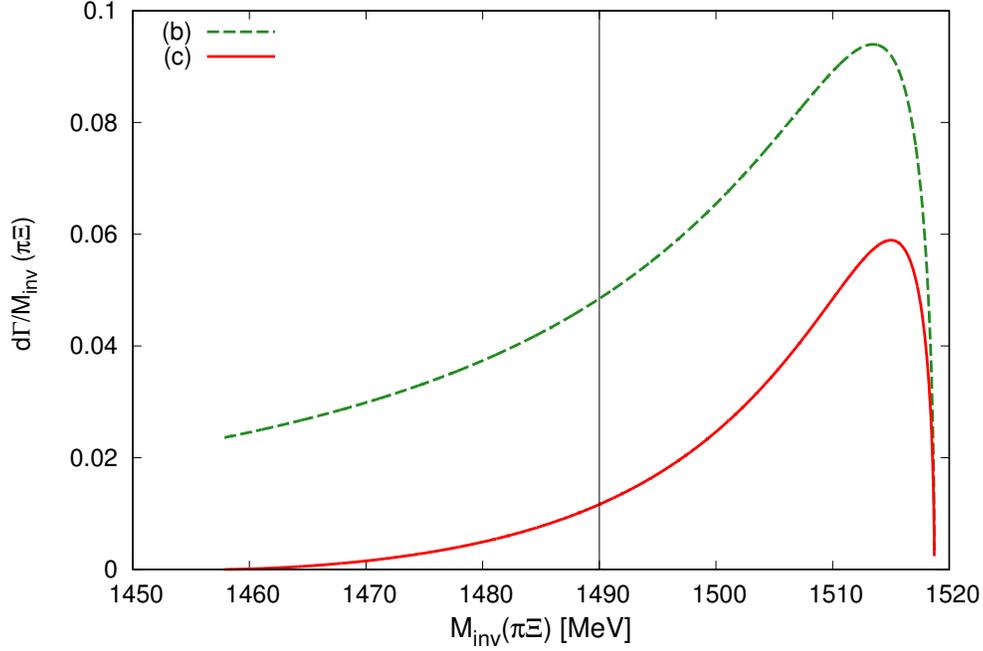}
\caption{ $d \Gamma_{\Omega^*}/ d M_{\rm inv}(\pi \Xi)$ of Eq.~(\ref{eq:dGamdM2}): (b) for the energy independent $\Xi^*$ width, and (c) for the energy dependent $\Xi^*$ width. The vertical line is the lower cut in $M_{\rm inv}(\pi \Xi)$ implemented in Ref.~\cite{belle2}.
}
\label{fig:4}
\end{figure}

Next we look into Eq.~(\ref{eq:dGamdM2}) and evaluate $d \Gamma_{\Omega^*}/ d M_{\rm inv}(\pi \Xi)$ through the mechanism of Fig.~\ref{fig:2}. We show the results in Fig.~\ref{fig:4}.
What we see in Fig.~\ref{fig:4} is $d \Gamma_{\Omega^*}/ d M_{\rm inv}(\pi \Xi)$ using Eq.~(\ref{eq:dGamdM2}), with the same coupling $g_{\bar K \Xi^*, {\rm con} }$ of Table~\ref{table:1} to facilitate the comparison, but taking $\Gamma_{\Xi^*}$ constant (option (b)) and $\Gamma_{\Xi^*}$ energy dependent (option (c)). We observe that the strength in the case of the energy dependent width is smaller than in the other cases, leading to a smaller $\Gamma_{\Omega^* \to \pi \bar K \Xi}$ width, in agreement with what was found in Eqs.~(\ref{eq:results.Gam}) from the observation of the shape of $|T|^2$. If we integrate the mass distribution of option (c) over $M_{\rm inv}(\pi \Xi)$, with or without the cut of Belle~\cite{belle2} of Eq.~(\ref{eq:belle.cut}), we find
\begin{equation}
\Gamma_{\Omega^* \to \pi \bar K \Xi} = 1.10~{\rm MeV},
\label{eq:result2.Gam}
\end{equation}
\begin{equation}
 \Gamma_{\Omega^* \to \pi \bar K \Xi {\rm (cut)}} = 0.98~{\rm MeV}.
\label{eq:result2.Gam.cut}
\end{equation}

We observe that the value $\Gamma_{\Omega^* \to \pi \bar K \Xi}$ from Eq.~(\ref{eq:result2.Gam}) is similar to the one obtained from Eqs.~(\ref{eq:results.Gam}) 
substituting  $ \Gamma_{\Omega^*, {\rm con (Edep)}} -  \Gamma_{\Omega^*, {\rm non}} = 1.00$~MeV. The result obtained from Eq.~(\ref{eq:result2.Gam}) should be considered more accurate. The next thing to consider is to see which is the effect of the experimental cut of Ref.~\cite{belle2} to obtain the ratio of Eq.~(\ref{eq:Exp.ratio}). We see that it leads to about 10~\% reduction. Finally, if we take $\Gamma_{\Omega^* \to \pi \bar K \Xi, {\rm (cut)}}$ from Eq.~(\ref{eq:result2.Gam.cut}) and divide by the width $\Gamma_{\Omega^*, {\rm non}}$ of Eqs.~(\ref{eq:results.Gam}), we obtain
\begin{equation}
 \frac{ \Gamma_{\Omega^* \to \pi \bar K \Xi {\rm (cut)}}}{\Gamma_{\Omega^*, {\rm non}}} =12.9~\%,
\label{eq:result2.ratio}
\end{equation}
in good agreement with the estimate of Eq.~(\ref{eq:res.ratio}). This is the ratio that should be compared with the experimental one, and, as we can see the ratio obtained is close to the experimental boundary, but still a bit higher.
We should emphasize that the $\bar K \Xi^*$ is bound considering $\Xi^*$ as an elementary particle. It is only the width of the $\Xi^*$, and its related mass distribution, what renders $\Gamma_{\Omega^* \to \pi \bar K \Xi}$ finite, but given the small width of the $\Xi^*$ one should expect small values of this width, as is the case here. There are other examples of that in hadron physics. Indeed, the $\Lambda(1520)$ appears as a dynamically generated resonance from the $\pi \Sigma^*$, $K \Xi^*$ channels in $s$-wave, but it mostly decays into $\bar K N$, $\pi \Sigma$ in $d$-wave~\cite{sarcarora}.

We should recall that with the molecular picture that we have studied, values of $\Gamma_{\Omega^* \to \pi \bar K \Xi}$ as we have obtained are unavoidable, and it is not possible to get smaller values while being consistent with the mass and full width of the $\Omega^*$. 
This is, however, using values of $\alpha$, $\beta$ of the same sign. The global sign does not matter, however, things could be different assuming a relative negative sign between $\alpha$ and $\beta$.\footnote{We thank J.~J.~Xie for calling our attention to this point.}
In view of this, we take new values of $\alpha$, $\beta$ of opposite sign and look again for acceptable solutions. It is possible to reduce the ratio of Eq.~(\ref{eq:result2.ratio}) with many solutions, but not drastically.
In view of this, we show the result with just one set of parameters.
Provided the experimental data are improved in the future, a best fit to determine the optimal parameters would be most advisable.

We find a reasonable solution with 
\begin{equation}
 q_{\rm max} = 735~{\rm MeV}; ~~ \alpha =-8.7 \times 10^{-8}~{\rm MeV}^{-3};~~
 \beta =18.3 \times 10^{-8}~{\rm MeV}^{-3}  ,
\label{eq:param.set.new}
\end{equation}
and we summarize the result in analogy to what was done before in Eq.~(\ref{eq:results.Gam}),
\begin{eqnarray} \label{eq:results.Gam_new}
&{\rm (a)}&~\Gamma_{\Omega^*, {\rm non}} = 7.3~{\rm MeV}, \nonumber\\
&{\rm (c)}&~\Gamma_{\Omega^*, {\rm con (Edep)}} = 8.1 ~{\rm MeV},\nonumber
\end{eqnarray}
\begin{equation}
~M_{\Omega^*} = 2012.7~{\rm MeV}, \nonumber 
\end{equation}
\begin{eqnarray}
\frac{\Gamma_{\Omega^*, {\rm con (Edep)}} - \Gamma_{\Omega^*, {\rm non}}}{\Gamma_{\Omega^*, {\rm non}}} = 10.9~\%,
\label{eq:res.ratio2}
\end{eqnarray}
which is already in agreement with experiment. The couplings slightly change with respect to those obtained before. They are summarized in Table~\ref{table:2}.
If we use new $|g_{i, {\rm conv}}|$ from Table~\ref{table:2} for the $\bar K \Xi^*$ channel and reevaluate Eq.~(\ref{eq:dGamdM2}), we get a ratio
\begin{equation}
 \frac{ \Gamma_{\Omega^* \to \pi \bar K \Xi {\rm (cut)}}}{\Gamma_{\Omega^*, {\rm non}}} =11~\%,
\label{eq:result2.ratio.new}
\end{equation}
in agreement with Eq.~(\ref{eq:res.ratio2}) and experiment.
We also see that the new solution gives a bit more weight to the $\eta \Omega$ component and a bit less to the $\bar K \Xi^*$ one. Together, they tell us that one has about 82~\% probability between the $\bar K \Xi^*$ and $\eta \Omega$ bound channels, stressing the molecular nature of the state.

\begin{table}[tb!]
\caption{\label{table:2} 
Same as Table~\ref{table:1} but with the parameters of Eq.~(\ref{eq:param.set.new}).
}
\begin{tabular}{cccc}
\hline\hline
& $\bar{K} \Xi^*$ (2027) & $\eta \Omega $ (2220) & $\bar{K} \Xi$ (1812)  \\
\hline
$g_i$ & $1.86 - i0.02$  &~~~$3.52-i0.46$ &~~~$-0.42 +i0.12$  \\
$|\tilde g_i|$ & 1.75   & 3.36   & 0.41   \\
$|g_{i, {\rm conv}}|$ & 1.72  & 3.30  &  0.41 \\
wf$_{ i}$($g_i G_i$) &~~~$-34.05 - i1.10 $  &~~~$-30.66 + i 3.67$  & $-$  \\
$-g_{i}^2 \frac{\partial G_i}{\partial \sqrt{s}}$ &~~~$0.57 + i0.10 $ & ~~~$0.25 - i0.06 $ & $-$  \\
\hline\hline
\end{tabular}
\end{table}

Coming back to the work of Ref.~\cite{pavao}, we summarize here the novelties of the present work which make the results compatible with the new Belle experimental threshold. We discussed that the consideration of the energy dependent width in the convolution of the $\Xi^*$ was a factor that drastically changes the ratio of Eq.~(\ref{eq:Exp.ratio}). On the other hand, the new choice of parameters allows for solutions compatible with this observable and in particular, the change of sign of the $\alpha$ and $\beta$ parameters is also new with respect to Ref.~\cite{pavao} and helped a bit improving the agreement with experiment. 
One might think that by obtaining this solution we might have made the $\bar K \Xi^*$ component much smaller, but, as seen in Table~\ref{table:2}, this is not the case and this component is still the most important one in the resulting state. We should not forget that without the $\bar K \Xi$ channel one can get a molecular state. Indeed, if we remove this channel we get a bound state of the other two channels with a mass $M_{\Omega^*, {\rm con (Edep)}} = 2012.5$~ MeV with a cut off $q_{\rm max} = 768$~MeV. It has a width $\Gamma_{\Omega^*, {\rm con (Eind)}} = 1.2$~MeV, which is fairly in
agreement with the value of Eq.~(\ref{eq:result2.Gam}) and the subtraction of the values (c) and (a) of Eq.~(\ref{eq:results.Gam}). 
The introduction of the $\bar K \Xi$ channel introduces an extra decay width but does not change qualitatively the nature of the state that one is obtaining.

We would like to discuss another issue at this point. The $\eta \Omega$ channel is bound by about 200~MeV with respect to the $\Omega(2012)$, or equivalently about 200~MeV above the $\bar K \Xi^*$ threshold. 
One may wonder whether one should not use higher order Lagrangians to deal with this channel. The contribution from higher order Lagrangians, or possible counterterms in effective theories, in the chiral unitary approach using lowest order Lagrangians is taken into account by means of a cut off or a subtraction constant in the regulation of the loops, which is finally fitted to some experimental value. We have used a same cut off for the $\bar K \Xi^*$ and $\eta \Omega$ channels. One may wonder if one should not use two different ones. In order to exploit this freedom, but keeping within what is called natural size of the cut off in Ref.~\cite{ollerulf}, of the order of 1 GeV, we see what results we obtain by making moderate changes in the cut off for the $\eta \Omega$ channel. Note that changes in the $\bar K \Xi$ channel are unnecessary, since a change in the cut off can be approximately accounted for by a change in the $\alpha$ or $\beta$ parameters.  

We find an acceptable solution with the set of parameters
\begin{equation}
 q_{\rm max} = 775~{\rm MeV};~~ q_{\rm max} (\eta \Omega) = 710~{\rm MeV}; 
~~ \alpha =-8.7 \times 10^{-8}~{\rm MeV}^{-3};~~
 \beta =18.3 \times 10^{-8}~{\rm MeV}^{-3}  ,
\label{eq:param.set.new_etaOmega1}
\end{equation}
leading to
\begin{eqnarray} 
&{\rm (a)}&~\Gamma_{\Omega^*, {\rm non}} = 7.7~{\rm MeV}, \nonumber\\
&{\rm (c)}&~\Gamma_{\Omega^*, {\rm con (Edep)}} = 8.5 ~{\rm MeV},\nonumber
\label{eq:results.Gam_new_etaOmega1}
\end{eqnarray}
\begin{eqnarray}
~M_{\Omega^*} = 2012.7~{\rm MeV}, \nonumber
\end{eqnarray}
\begin{eqnarray}
\frac{\Gamma_{\Omega^*, {\rm con (Edep)}} - \Gamma_{\Omega^*, {\rm non}}}{\Gamma_{\Omega^*, {\rm non}}} = 10.4~\%.
\label{eq:res.ratio_etaOmega1}
\end{eqnarray}

\begin{table}[tb!]
\caption{\label{table:3} 
Same as Table~\ref{table:1} but with the parameters of Eq.~(\ref{eq:param.set.new_etaOmega1}).
}
\begin{tabular}{cccc}
\hline\hline
& $\bar{K} \Xi^*$ (2027) & $\eta \Omega $ (2220) & $\bar{K} \Xi$ (1812)  \\
\hline
$g_i$ & $1.79 + i0.02$  &~~~$3.79-i0.53$ &~~~$-0.44 +i0.14$  \\
$|\tilde g_i|$ & 1.68   &  3.60   &  0.43   \\
$|g_{i, {\rm conv}}|$ & 1.65  &  3.56  &  0.43  \\
wf$_{ i}$($g_i G_i$) &~~~$-34.50 - i1.80 $  &~~~$-31.25 + i 3.96$  & $-$  \\
$-g_{i}^2 \frac{\partial G_i}{\partial \sqrt{s}}$ &~~~$0.53 + i0.12 $ & ~~~$0.28 - i0.07 $ & $-$  \\
\hline\hline
\end{tabular}
\end{table}

Another possible solution is given by
\begin{equation}
 q_{\rm max} = 735~{\rm MeV};~~ q_{\rm max} (\eta \Omega) = 750~{\rm MeV}; 
~~ \alpha =-11.0 \times 10^{-8}~{\rm MeV}^{-3};~~
 \beta =20.0 \times 10^{-8}~{\rm MeV}^{-3}  ,
\label{eq:param.set.new_etaOmega2}
\end{equation}
which leads to
\begin{eqnarray}
&{\rm (a)}&~\Gamma_{\Omega^*, {\rm non}} = 8.2~{\rm MeV}, \nonumber\\
&{\rm (c)}&~\Gamma_{\Omega^*, {\rm con (Edep)}} = 9.1 ~{\rm MeV},\nonumber
 \label{eq:results.Gam_new_etaOmega2}
\end{eqnarray}
\begin{eqnarray}
~M_{\Omega^*} = 2012.6~{\rm MeV}, \nonumber 
\end{eqnarray}
\begin{eqnarray}
\frac{\Gamma_{\Omega^*, {\rm con (Edep)}} - \Gamma_{\Omega^*, {\rm non}}}{\Gamma_{\Omega^*, {\rm non}}} = 10.9~\%.
\label{eq:res.ratio_etaOmega2}
\end{eqnarray}
We summarize the information obtained on couplings, wave functions and probabilities in Tables~\ref{table:3} and~\ref{table:4}.

\begin{table}[tb!]
\caption{\label{table:4} 
Same as Table~\ref{table:1} but with the parameters of Eq.~(\ref{eq:param.set.new_etaOmega2}).
}
\begin{tabular}{cccc}
\hline\hline
& $\bar{K} \Xi^*$ (2027) & $\eta \Omega $ (2220) & $\bar{K} \Xi$ (1812)  \\
\hline
$g_i$ & $1.88 + i0.04$  &~~~$3.55-i0.67$ &~~~$-0.42 +i0.22$  \\
$|\tilde g_i|$ & 1.77   &  3.42   &  0.44   \\
$|g_{i, {\rm conv}}|$ & 1.75  &  3.38  &  0.45  \\
wf$_{ i}$($g_i G_i$) &~~~$-34.37 - i2.42 $  &~~~$-31.99 + i 5.63$  & $-$  \\
$-g_{i}^2 \frac{\partial G_i}{\partial \sqrt{s}}$ &~~~$0.57 + i0.16 $ & ~~~$0.26 - i0.09 $ & $-$  \\
\hline\hline
\end{tabular}
\end{table}

These are just two possible solutions and one could find more, but the important message is that, even accepting this new freedom, once the experimental constraints of Eqs.~(\ref{eq:Exp.mass}),(\ref{eq:Exp.width}),(\ref{eq:Exp.ratio}) are imposed, the results on the couplings, the wave functions at the origin and the probabilities of the closed channels are very stable, stressing the molecular nature of the state. The differences between the results in Tables~\ref{table:2},\ref{table:3},\ref{table:4} give us an idea of the uncertainties that we have.

  We can add further information to this respect. Simultaneously and independently, a work on this issue was done in Ref.~\cite{xiegeng2} using the same formalism as here and Ref.~\cite{pavao}. The works are complementary in the information they provide, and share the same conclusions. In Ref.~\cite{xiegeng2} other uncertainties are investigated and different best fits are done to the data to get the optimal parameters. The values of the parameters include the solutions that we have reported but they explore a wider range. They keep $q_{\rm max}$ equal for all the channels and they find that there is freedom choosing different values of $q_{\rm max}$ bigger than 720 MeV and adjusting the $\alpha$ and $\beta$ parameters to agree with the experimental data in each case. Their results are summarized in Tables II and III of that reference. By varying $q_{\rm max}$ from 735~MeV to 900~Mev the masses are always 2012~MeV but the total widths change from 8.3 to 6.4~MeV, still within the experimental range. The values for $R$ are also in the range of 11--7~\%. The conclusions of the paper are the same as ours: the data are compatible with the molecular picture and the results are rather stable with different sets of parameters within natural size. Future better precision of the experimental data will allow also to be more refined in the fits and provide more precise output.

\section{Summary and conclusions}\label{conclusion}
We have made a thorough study of the viability of the molecular picture for the $\Omega(2012)$ state in view of the boundary found in the Belle experiment for the $\Omega^*$ width going to $\pi \bar K \Xi$, as a signal of the $\bar K \Xi^*$ component. The study is rather complete and contains the $\bar K \Xi^*$, $\eta \Omega$ and $\bar K \Xi$ states as coupled channels in a unitary approach.
The transition potential between $\bar K \Xi^*$ and $\eta \Omega$ are taken from the chiral Lagrangians but the transition potentials from $\bar K \Xi^*$, $\eta \Omega$ in $s$-wave to the $\bar K \Xi$ in $d$-wave are taken as free parameters. Together with a cut off to regularize the loops, this provides three unknown magnitudes in the theory which are fitted to the data to reproduce the mass, width and partial decay width of the $\Omega(2012)$ to $\pi \bar K \Xi$. We find an acceptable solution in terms of natural values for the parameters which reproduce fairly well the experimental data. Here, the main conclusion is the compatibility of the molecular picture with present data. Yet, we also observe that it is not possible to obtain ratios of $\Gamma_{\Omega^* \to \pi \bar K \Xi}/\Gamma_{\Omega^* \to \bar K \Xi}$ smaller than about 13~\% without spoiling the agreement with mass and width of the $\Omega^*(2010)$ if the $\alpha, \beta$ parameters are taken of the same sign. But the ratio is improved for sets with $\alpha, \beta$ of opposite sign, leading to values compatible with experiment.
A more precise measurement, providing the $\Gamma_{\Omega^* \to \pi \bar K \Xi}/\Gamma_{\Omega^* \to \bar K \Xi}$ ratio, or finding a much stringent upper bound than the present one, will be determining to settle the issue of the possible molecular picture for the $\Omega(2012)$ state. We should note that the molecular structure of this state is rather peculiar, in the sense that it corresponds to mostly a $\bar K \Xi^*$ bound state, however, it requires the interaction with the $\eta \Omega $ channel to bind, while neither the $\bar K \Xi^*$ nor the $\eta \Omega$ states would be bound by themselves.

\section{ACKNOWLEDGEMENT}
The work of N. I. was partly supported by JSPS Overseas Research Fellowships and JSPS KAKENHI Grant Number JP19K14709.
G. T. acknowledges the support of PASPA-DGAPA, UNAM for a sabbatical leave. 
This work is partly supported by the Spanish Ministerio de Economia y
Competitividad and European FEDER funds under Contracts
No. FIS2017-84038-C2-1-P B and No. FIS2017-84038-C2-2-P B.
This project has received funding from the European Union’s Horizon 2020 research and innovation programme under grant agreement No 824093 for the **STRONG-2020 project.

\end{document}